%% file: main.tex
\pdfoutput=1
\documentclass{article}

\PassOptionsToPackage{numbers, compress}{natbib}

 \usepackage[dblblindworkshop, final]{neurips_2026}

\workshoptitle{AI for Stochastic Dynamics}

\usepackage[utf8]{inputenc} %
\usepackage[T1]{fontenc}    %
\usepackage{hyperref}       %
\usepackage{url}            %
\usepackage{booktabs}       %
\usepackage{multirow}       %
\usepackage{amsfonts}       %
\usepackage{amsmath}        %
\usepackage{graphicx}       %
\usepackage{microtype}      %
\usepackage{xcolor}         %

\input{notation.tex}        %

\title{Only Linear Constraints Survive Coarse-Graining: Evaluating Physics-Constrained Neural
  Operators on Stochastically Forced Turbulence}

\author{%
  Michael Groom \\
  CSIRO Environment, Hobart, Australia \\
  \texttt{Michael.Groom@csiro.au} \\
  \And
  Rafael Oliveira \\
  CSIRO Technology, Sydney, Australia \\
  \texttt{Rafael.Dossantosdeoliveira@csiro.au} \\
}

\begin{document}

\maketitle

\begin{abstract}
  A physics-informed loss adds a discretised PDE residual to the data term, but is only strictly valid if the equation(s) being enforced are closed on the fields being fitted. An example of where this is not valid is on coarse-grained, stochastically forced data: nonlinear terms in the PDE generally do not commute with the filter, so the coarse-grained fields do not satisfy the original equations, and attempting to enforce these equations as constraints by driving their residual to zero encodes an implicit closure into the learned operator. Under normalised, translation-invariant filtering on a periodic domain, constant-coefficient linear constraints remain valid on the filtered grid. For the 2D incompressible Navier--Stokes equations, two such constraints that remain valid are continuity and global momentum balance. We enforce these exactly by applying a closed-form projection on the output of a Fourier neural operator (FNO) at $O(N \log N)$ cost. On two-dimensional isotropic turbulence, truncated so that its forcing lies entirely beyond the cutoff wavenumber, the projection achieves a continuity error of $\max|\diver \vec{u}|=5.0\times10^{-7}$
  against $0.51$ for a plain FNO and $0.17$ for a physics-informed FNO, and a global momentum balance error of $|\langle \boldsymbol{u} \rangle_\Omega|=8.6\times10^{-11}$ against $7.5\times10^{-4}$ and $8.3\times10^{-4}$. The projection only costs an extra 2\% of training time, and takes the fraction of rollouts whose energy remains bounded at $448$ steps from $0.18$ to $0.87$. On resolved Kolmogorov flow, test-time optimisation costs approximately $3{,}300\times$ as much per trajectory as the projection, while leaving maximum divergence three orders of magnitude higher. Nonlinear constraints can also be enforced once their unclosed terms are modelled: closing the (quadratic) global energy balance with a constant subgrid flux, fitted from the coarse training data alone, removes the $20.9$\% energy deficit that enforcing the unclosed balance produces and keeps every free-running rollout bounded over $2{,}000$ steps.
\end{abstract}

\input{sections/01_introduction.tex}

\input{sections/05_results.tex}
\input{sections/06_discussion.tex}
\input{sections/07_conclusion.tex}

\begin{ack}
The authors wish to acknowledge Xuesong Wang for providing the initial training and evaluation pipeline for the FNO models used in this paper. We thank the anonymous reviewers and the area chair for their constructive comments.
\end{ack}

\appendix

\input{sections/02_filtering.tex}
\input{sections/03_projection.tex}
\input{sections/04_setup.tex}

\input{appendix/I_closure.tex}

\input{appendix/E_tto.tex}
\input{appendix/G_reproduction.tex}

\end{document}

%% file: notation.tex
\newcommand*{\R}{\mathbb{R}}                    %
\newcommand*{\domain}{\Omega}                   %

\renewcommand{\vec}[1]{\boldsymbol{#1}}         %
\newcommand*{\mat}[1]{\mathbf{#1}}              %
\newcommand*{\transpose}{\mathsf{T}}            %

\newcommand*{\diver}{\nabla\cdot}               %

\newcommand*{\pdeop}{\mathcal{L}}               %
\newcommand*{\locop}{\mathcal{D}}               %
\newcommand*{\bcop}{\mathcal{B}}                %
\newcommand*{\globop}{\mathcal{I}}               %

\newcommand*{\filt}[1]{\overline{#1}}           %
\newcommand*{\sgs}{\vec{\tau}}                  %

\newcommand*{\nnop}{\mathcal{G}}                %
\newcommand*{\parameters}{\theta}               %
\newcommand*{\proj}{\mathcal{P}}                %

%% file: sections/01_introduction.tex
\section{Introduction}
\label{sec:introduction}

Learning an evolution operator from finitely many samples is intrinsically underdetermined: many operators can interpolate the observed data while representing fundamentally different dynamics away from the training set. Known physical laws provide information that the observations alone cannot. They constrain the admissible states and trajectories of the system independently of the particular initial conditions, sampling interval, or time horizon represented in the data, and therefore provide a principled means of constraining the model. This matters not only for extrapolation, but also because predictions are commonly used by downstream models and analyses that implicitly assume that those physical constraints are satisfied. Neural operators can incorporate this information either softly, through penalties added to the training objective, or exactly, by restricting the model architecture or its outputs to the physically admissible set. Physics-informed losses provide one widely used example of the former approach: a discretised partial differential equation (PDE) residual is added to the data-fitting objective, as in physics-informed neural networks \citep[PINNs;][]{raissi2019pinn} and their operator-learning counterpart, the physics-informed neural operator \citep[PINO;][]{li2021pino}. Such a residual, however, represents a valid physical constraint only when the governing equations are closed on the fields being fitted. When this condition is satisfied, the PDE supplies additional information without requiring labelled solutions and can often be evaluated inexpensively throughout the entire domain. When it is not, enforcing the residual does not make the model more physical: it imposes an equation that the predicted fields cannot, in general, satisfy. The data-fitting and physics objectives then become incompatible, and training can only select a compromise between them \citep{rohrhofer2023pareto}.

An example of where physics-informed training is not valid is on coarse-grained, stochastically forced data. This is because a
coarse-grained flow field, for example a macroscopic observation of a multiscale physical system, is a filtered image of a finer resolution state. Nonlinear terms in the equations governing that state generally do not commute with the filter, so the coarse-grained fields do not satisfy the
original equations. This is the classical closure problem of large-eddy simulation \citep[LES,][]{pope2000turbulent,sagaut2006les}. Recent work on this problem in the machine learning literature either learns the unresolved term from direct numerical simulation \citep{guan2023physicsconstrained}, inserts a classical subgrid model into a physics-informed loss \citep{zhao2025lesnets}, or evaluates the residual on a fine grid \citep{wang2024coarsegraining}. These all either rely on data not available from the coarse field alone, or inherit the shortcomings of classical LES models. A PDE-residual loss
evaluated solely on the coarse field makes the same approximation as LES, but left implicit. Attempting to enforce the
original equations as constraints by driving their residual to zero encodes an
unintended closure into the learned operator, selected by the optimiser rather than the modeller.

Linear constraints that commute with the filter survive coarse-graining exactly. On a periodic
domain, a normalised, translation-invariant filter commutes with constant-coefficient
differential operators and preserves domain integrals; linearity alone does not guarantee
this for arbitrary operators or boundary conditions. The criterion is therefore commutation with the filter rather than linearity as such; linearity is what then lets a surviving constraint be enforced exactly by projection. In this paper we use the 2D incompressible Navier--Stokes equations as a motivating example, for which two
linear constraints that remain valid are continuity and global momentum balance obtained by
integrating the momentum equation over the domain. By contrast, the pointwise momentum equation and global energy balance
are both nonlinear, therefore neither remains valid to enforce as a constraint after filtering without some form of learned/modelled closure. Here we do not attempt to close the unresolved terms in the pointwise momentum equation. Instead we show, on stochastically forced two-dimensional turbulence, which constraints remain valid after filtering and what happens when a neural operator is trained as if they all were. The motivating setting is neural operators that see only coarse-grained observations (for example, geophysical fields in weather/climate prediction); the experiments here are a controlled proxy for that case.

In addition to remaining valid after filtering, another advantage of linear constraints is that they can often be imposed exactly via the solution of a small,
equality-constrained, least-squares problem \citep{chen2024kkthpinn,min2024hardnet}. Furthermore, on a
periodic grid that closed form solution is diagonal in Fourier space, so it only requires one forward and one
inverse transform; no linear system needs to be solved. We therefore enforce continuity and global momentum balance cheaply and exactly by
applying this projection to the output of an FNO at
$O(N \log N)$ cost. The projection has no parameters, is differentiable, and composes with the model in end to
end training. Unlike a residual penalty, it requires no coefficient to weight against the data term in the loss.
In situations where it is valid to do so, we can also impose global energy balance, a single quadratic
equation, using a Gauss--Newton projection applied after the linear step \citep{utkarsh2025pcfm}.
The update is a rank-one correction whose direction is divergence-free wherever the field is and has
no domain mean, so continuity and the momentum balance remain enforced. On coarse data the energy
balance is not closed, so imposing it exactly requires a model for the unresolved energy flux across
the cutoff. We test the simplest such model: a constant, fitted from the coarse training data alone,
which shifts the target of the constraint by one scalar and turns a mis-specified constraint into a
valid one.

We consider two test cases to both demonstrate the effects of enforcing unclosed constraints, as well as the effectiveness of exactly enforcing closed ones. The first is the (resolved) Kolmogorov flow benchmark from
\citet{li2021pino}, where the equations are closed up to
discretisation error. The second is two-dimensional isotropic turbulence
\citep{boffetta2012twodimensional}, simulated using a pseudo-spectral solver at $1024^{2}$ and truncated to $64^{2}$ by a sharp
spectral cutoff placed \emph{below} the forcing band, so that the forcing lies entirely beyond the
cutoff wavenumber. %
On this unresolved flow the forcing is invisible to the learner as it is entirely subgrid, so the energy dissipated in the resolved band is supplied entirely by nonlinear transfer from unresolved scales. The map from one coarse snapshot to the next is therefore driven by scales that are not in
the input and are not deterministic functions of it: many fine states are compatible with the same coarse field, and they evolve differently. A deterministic operator trained under squared error can still learn the conditional mean of this stochastic transition. The constraints that remain valid for that mean are again the linear ones, because the expectation is linear as well.

\paragraph{Contributions.}
\begin{enumerate}
  \setlength{\itemsep}{1pt}
  \item A filtering criterion that separates the constraints which remain valid after
  coarse-graining from those which do not: a residual loss on filtered data is an implicit
  closure model.
  \item A spectral projection that enforces continuity and global momentum balance \emph{jointly} on an FNO, in closed form, at $O(N \log N)$ cost, together with a Gauss--Newton step for the global energy balance that leaves those two constraints intact.
  \item A comparison of that projection against the PINO residual, with and without
  test-time optimisation, on a resolved control and an unresolved case, to demonstrate the effects of mis-specified physical constraints.
  \item A proof of concept that a nonlinear global constraint can be made valid on coarse data by
  closing it: a constant subgrid energy flux, one scalar fitted from the coarse training pairs
  alone, removes the energy bias that enforcing the unclosed balance produces.
\end{enumerate}

\paragraph{Relation to prior work.} Closed-form
projection onto linear equality constraints as a non-trainable neural network output layer was introduced in
KKT-hPINN \citep{chen2024kkthpinn} and HardNet \citep{min2024hardnet}, is treated in generality by
\citet{grontas2026pinet}, and appears in iterative or implicit form in DC3
\citep{donti2021dc3} and in differentiable-optimisation layers
\citep{negiar2023differentiable,chalapathi2024scaling}. A spectral divergence-free projection
appended to an FNO and trained end to end on two-dimensional Navier--Stokes has been published in 
\citep{duruisseaux2024hard,xu2025pcno,li2026projectgenerate}; what we add to that layer is the
global momentum balance constraint in the same step. Quadratic integral constraints have been
projected in PINNs \citep{baez2025pinnproj} and imposed on FNOs by a learnable correction
\citep{liu2025adaptivecorrection}; the energy step here is parameter-free, applied after the
spectral projection, and constructed so that it does not undo continuity or the momentum balance.
Hard constraints by reparameterisation are an alternative route to the same exactness
\citep{richterpowell2022neural,beucler2021enforcing,mohan2023embedding,liu2024clawno}, and
projections onto global conservation laws are established for PINNs and operators alike
\citep{hansen2023probconserv,baez2025pinnproj,liu2025adaptivecorrection,utkarsh2025pcfm}. The
closest quantitative comparison of hard vs. soft constraints is \citet{baez2025pinnproj}, who report three to
four orders of magnitude lower conservation error than a soft penalty.
\citet{harder2023hardconstrained} enforce a linear coarse-to-fine mass balance in climate downscaling: a related constraint that also survives coarse-graining, but not the dynamical constraints considered here.
None of this work uses a filtering argument to decide which constraints to enforce.
The comparisons below use one architecture, one optimiser and a common schedule across methods within each flow, based on the training and evaluation pipeline of \citet{wang2026mswt}, so that differences are not artefacts of weak or unmatched baselines \citep{mcgreivy2024weak}.

%% file: sections/05_results.tex
\section{Results}
\label{sec:results}

The filtering criterion, the projection, and the description of two flows are given in
\autoref{sec:filtering}--\autoref{sec:setup}. For each experiment the eight methods share one architecture, one optimiser and
one schedule, with five seeds each (\autoref{app:protocol}). These are a plain FNO, an FNO with the projection applied (continuity, continuity with the momentum balance, and both with the
energy balance), a PINO with the full continuity + momentum residual or continuity only; and each PINO method with additional test-time
optimisation (TTO) during the rollout. \autoref{sec:results-probe} adds one further model on E2: the
energy-balance projection with a fitted closure, trained under the same protocol. For both experiments we compare accuracy, constraint satisfaction, stability over long rollout horizons and computational cost.

\begin{figure}[t]
  \centering
  \includegraphics[width=1.0\linewidth]{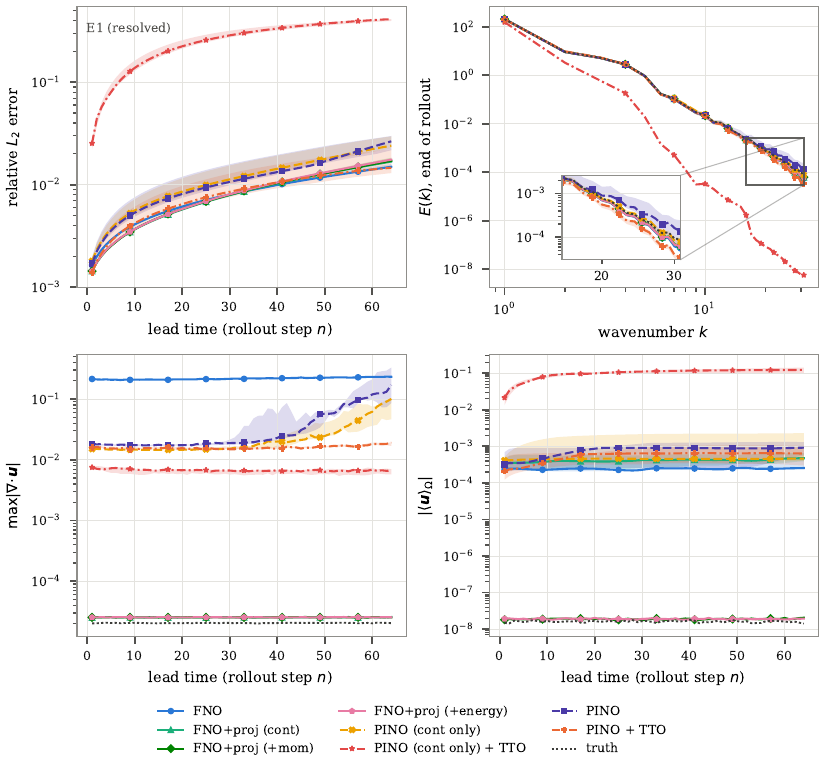}
  \caption{E1 (the resolved case): relative $L_{2}$ error, the step-$64$ energy spectrum,
  maximum divergence and domain-mean velocity. Curves and shading show the median and
  interquartile range over five seeds, each averaged over $200$ test rollouts ($20$ for TTO).}
  \label{fig:e1-rollout}
\end{figure}

\begin{figure}[t]
  \centering
  \includegraphics[width=1.0\linewidth]{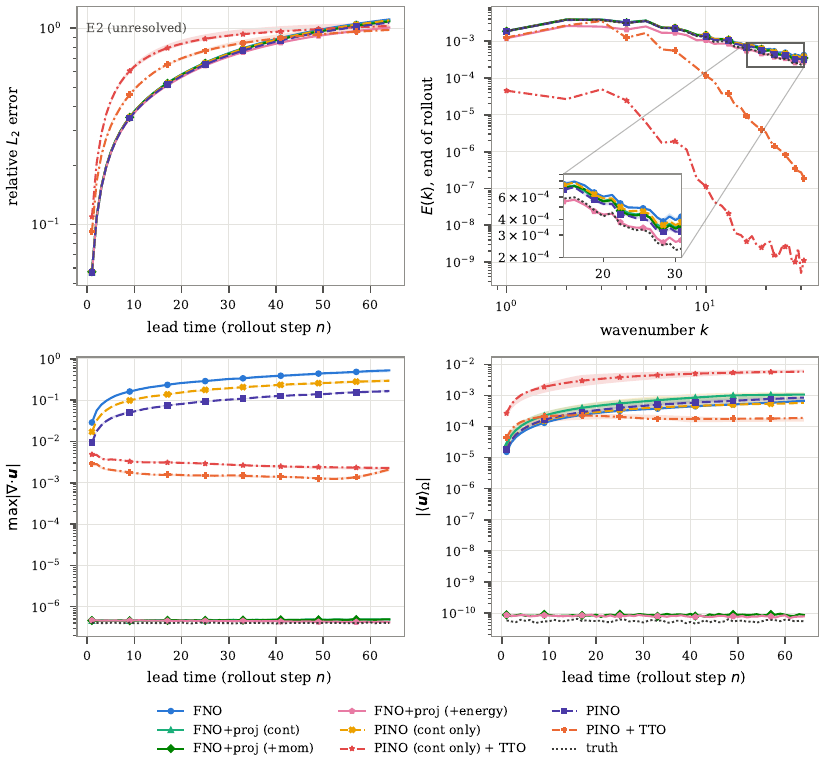}
  \caption{As in \autoref{fig:e1-rollout} but for E2 (the unresolved case), averaged over each seed's $42$ test rollouts.}
  \label{fig:e2-rollout}
\end{figure}

\begin{table}[t]
  \caption{Relative $L_2$ error at the final step of the $64$-step rollout against the ground truth, constraint
  satisfaction over the same rollout, and training and inference costs, for both flows. Entries are means over five seeds
  with the seed standard deviation. $\max|\diver \vec{u}|$ is taken at the worst step of the
  rollout and $|\langle \boldsymbol{u} \rangle_\Omega|$ at the final step. Cost columns are relative to the FNO for the same flow. Daggers mark E2 step-$64$ errors that are lower because the model has lost energy, not because it tracks the flow (\autoref{sec:results-accuracy}).}
  \label{tab:main}
  \centering
  \footnotesize
  \setlength{\tabcolsep}{2pt}
  \input{figures/table1_body.tex}
\end{table}

\subsection{Accuracy}
\label{sec:results-accuracy}

\autoref{fig:e1-rollout} shows the E1 rollout errors; \autoref{tab:main} gives their final values. On E1, the Kolmogorov flow where the momentum equation closes, projecting the two linear constraints gives a small and consistent
gain at short lead times: $12.3 \pm 2.5$\% lower one-step error than the plain FNO, better
on five seeds of five, and $11.5$\% lower at step $16$. By the final step of the $64$-step
rollout the gain has gone: the projected model predictions are $6.9 \pm 16.0$\% worse than the
FNO, although that difference is well inside the seed spread. Adding the residual loss on the same flow is a
different story. PINO is $45.8$\% worse than the FNO at step $16$ and $61.9 \pm 37.3$\% worse
at step $64$, and PINO (cont only) is similar at $58.6$\%. On a
resolved flow, where adding the PDE residual would be expected to help, it does not (at least when only applied during the training stage). Meanwhile, the exact satisfaction of the constraints that the projection provides is, at worst, free in terms of accuracy. Further optimisation of the full residual at test time (PINO+TTO) results in $10.1 \pm 12.6$\% lower error relative to the plain FNO at step $1$, a gain inside the seed spread and no better than the projection's,
and $7.6$\% worse at step $64$. Optimising continuity
alone gives a final relative $L_2$ error of $0.425$, $26\times$ worse than the FNO, for the reason given in \autoref{app:tto-cont}.

\autoref{fig:e2-rollout} and \autoref{tab:main} give the corresponding E2 results. On E2, the stochastically forced turbulent flow where the momentum equation does not close, the accuracy comparison is
unchanged at short lead times. At step $1$, the relative $L_2$ errors of the FNO, the three projected models and both PINOs span $0.0569$ to $0.0574$, and
compared seed by seed none differs from the FNO by more than $0.5$\%.
By step $64$ ($25.6$ time units) every method has a relative error above $1.0$, worse than
predicting zero everywhere. Step $64$ is eight $e$-folding times of the coarse field ($3.2$ time
units), so the flow has ``forgotten'' its initial condition.
The remaining differences, $-1.9$\% for the projection, $-3.9$\% for PINO and $-9.0$\%
when the energy balance is also imposed, therefore measure remaining energy, not agreement
with the ground truth: the lowest error amongst the PINOs (without TTO) and the projected FNOs is the energy balance projection, which removes the
most energy (\autoref{sec:results-probe}).
Further optimisation of the full residual at test time (PINO+TTO) gives a one-step error of
$0.093$ against $0.057$ for PINO, and $0.109$ when only continuity is optimised; at step $64$
those errors are $4.5$\% and $4.1$\% lower, again a difference in energy. Both TTO methods
lose energy from the first steps: in the free-running rollouts of \autoref{sec:results-stability},
the continuity-only TTO trajectories fall below a tenth of their initial energy at a median
step of $22$, so these step-$64$ errors describe a drained field.

\subsection{Constraint satisfaction}
\label{sec:results-exact}

The bottom two panels of \autoref{fig:e1-rollout} and \autoref{fig:e2-rollout} show constraint
satisfaction over the rollout, $\max|\diver \vec{u}|$ and $|\langle \boldsymbol{u} \rangle_\Omega|$, which are also listed in \autoref{tab:main}.
On E2, over the $64$-step rollout, the projected models maintain $\max|\diver \vec{u}|$ at
$5.0 \times 10^{-7}$ against the ground truth's $4.2 \times 10^{-7}$, contrasted with $0.51$ for the
FNO, $0.17$ for PINO and $0.30$ for PINO (cont only). Adding the momentum balance maintains the
domain-mean velocity at the final rollout step at $8.6 \times 10^{-11}$ against the ground truth's
$5.4 \times 10^{-11}$, whereas the FNO only manages $7.5 \times 10^{-4}$, PINO
$8.3 \times 10^{-4}$ and PINO (cont only) $9.0 \times 10^{-4}$. The projections therefore give six orders of
magnitude better constraint satisfaction on continuity and seven on the momentum balance. On E1, the projection likewise reaches the precision of the stored data: $\max|\diver \vec{u}|$ is $2.6 \times 10^{-5}$ against the ground truth's $2.1 \times 10^{-5}$ for the projection vs. $0.23$, $0.22$ and $0.096$ for the FNO, PINO and PINO (cont only), and $|\langle \boldsymbol{u} \rangle_\Omega|$ is $2.0 \times 10^{-8}$ against the ground truth's $1.4 \times 10^{-8}$ for the projection vs. $3.3 \times 10^{-4}$, $1.0 \times 10^{-3}$ and $1.1 \times 10^{-3}$ for the FNO, PINO and PINO (cont only).
Further optimisation of the full residual at test time (PINO+TTO) reduces $\max|\diver \vec{u}|$
to $2.4 \times 10^{-2}$ on E1 and $2.8 \times 10^{-3}$ on E2, below PINO but still two and a
half to four orders of magnitude above the projection. Optimising continuity alone gives
$8.0 \times 10^{-3}$ and $5.0 \times 10^{-3}$ on the same diagnostic, and increases the
domain-mean velocity by $100\times$ on E1 and $8\times$ on E2 (see \autoref{app:tto-cont}).

\subsection{Long-term stability}
\label{sec:results-stability}

The second panel of \autoref{fig:e1-rollout} and \autoref{fig:e2-rollout} shows the kinetic energy
spectrum at the final step of the $64$-step rollout.
On E1, the ground truth contains only
$0.0048$\% of its total energy above radial wavenumber $16$, and at step 64 of the rollout the FNO has
$0.0042$\% in that band; the projected FNO models and both PINOs follow the same spectrum.
Further optimisation of the full residual at test time (PINO+TTO) also follows that spectrum, whereas optimising continuity alone leaves only $0.00019$\% of energy above mode 16.
On E2, the
ground truth now contains $18.3$\% of its total energy in that same band. E2 uses $32$ rather
than $16$ modes per axis to extend the spectral convolution to smaller scales; this rectangular
mode selection differs from the radial band used for the diagnostic. At step 64 of the rollout the FNO, the three projected models
and both PINOs have $21.3$\% to $23.9$\% above mode 16, whereas PINO+TTO has only $0.20$\% of its energy above mode 16, and optimising continuity alone has $0.015$\%. The momentum equation does not close on this flow, therefore driving its residual to zero at test time removes energy from that wavenumber band rather than matching the truth.

To test long-term stability, free-running rollouts were conducted on both E1 and E2 out to 2,000 steps (i.e. beyond the horizon of the ground truth, which is available to 64 steps on E1 and 448 steps on E2). \autoref{fig:stability} reports the fraction of trajectories that remain bounded: defined as their kinetic energy staying within a factor of ten of its initial value, above or below
(\autoref{app:protocol}). On E1, every method is still at $1.00$ (i.e. 0\% of trajectories have left the band) at step $64$. At step $448$ the FNO and the three projected
models remain at $1.00$, PINO at $0.91$ and PINO (cont only) at $0.96$. At $2{,}000$ steps the
projected models remain at $0.99$ to $1.00$ and the FNO at $0.96$, while PINO has fallen to $0.42$ and PINO
(cont only) to $0.64$. Further optimisation of the full residual at test time (PINO+TTO) does not
prevent that fall ($0.52$), and nor does optimising continuity alone ($0.38$). On this flow, minimising the PDE residual costs stability at long rollout horizons, whereas the projection does not. On E2, the bounded fraction is also $1.00$ for every method at step $64$ except the two TTO methods, which lose energy rather than gain it: PINO+TTO is at $0.99$ and PINO (cont only)+TTO at $0.12$. By step $448$ it is $0.18$ for the
plain FNO, $0.71$ with continuity projected, $0.87$ with the momentum balance, $1.00$ with the
energy balance, $0.89$ for PINO and $0.40$ for PINO (cont only). PINO+TTO has fallen to $0.38$ and PINO (cont only)+TTO remains at $0.12$; neither method blows up, but both drain substantial amounts of energy from the flow and leave the band from below. At $2{,}000$ steps the FNO, the continuity-only projection and both PINOs are at $0.00$, the
momentum-balance projection at $0.12 \pm 0.28$ (one seed out of five has $0.62$ whereas the other four are none), the energy-balance projection at $0.00$, PINO+TTO at $0.20$
and PINO (cont only)+TTO at $0.12$. Again, the last three leave the band from below rather than above: they lose energy rather than gain it, with the energy balance projection losing it most slowly (its trajectories are the last to fall below $0.1\times$, from step $448$ onwards). The closed-energy model of \autoref{sec:results-probe} is also shown, and remains at $1.00$ throughout.

\subsection{Computational cost}
\label{sec:results-cost}

The last two columns of \autoref{tab:main} show training and inference costs for each method relative to the plain FNO. On E1, projecting the two linear constraints costs an additional $4$\% of training time, the residual loss an additional
$10$\%, and the energy projection $49$\%. On E2 the same overheads are $2$\%, $5$\% and $23$\%. At inference the projection costs $1.04\times$ and $1.07\times$ the FNO forward pass on E1 for the continuity projection or continuity plus momentum balance, and $1.34\times$ if the energy balance is included as well. On E2 the same costs are $1.04\times$, $1.05\times$ and $1.34\times$. PINO inference is a negligible additional cost, whereas further optimisation of the full residual at test time (PINO+TTO) is $3535\times$ the FNO on E1 and $2237\times$ on E2; TTO of continuity alone is $7312\times$ and $1388\times$.

\subsection{A simple energy balance closure}
\label{sec:results-probe}

The energy balance does not close on E2 because the resolved band receives energy via nonlinear
transfer across the cutoff (\autoref{sec:setup}). The unclosed energy-balance projection ends
the rollout with energy $20.9 \pm 0.1$\% below the ground truth. The truth's measured mean
residual of $1.28\times10^{-4}$ per step gives an estimated $22.7$\% deficit over $64$ steps relative
to its initial energy of $3.61\times10^{-2}$. Restoring that balance requires a closure energy flux, and we test the simplest one: a constant energy source $\bar{S}$, fitted as the mean
unclosed residual of the coarse training pairs, so that no fine-grid data is needed
(\autoref{app:closure}). On E2 the fit gives $\bar{S} = 3.20\times10^{-4}$ per unit time, and a
radial shell-flux measurement at $K=31$ agrees with it to within $7$\%. %

With $\bar{S}$ in the projector and nothing else changed, the bias disappears: the energy
error at step $64$ goes from $-20.9 \pm 0.1$\% to $-0.2 \pm 0.1$\%, and over $2{,}000$
free-running steps the closed model is the only one on E2 whose median energy remains near its
initial value ($0.71$ at step $2{,}000$), with all $210$ trajectories inside $[0.1, 10]\times$
their initial energy (\autoref{fig:stability}). Accuracy and satisfaction of the two linear constraints are unchanged.
The closure does not correct how that energy is distributed over wavenumbers: the fraction
above wavenumber $16$ still grows from the truth's $0.18$ to $0.50$ by step $2{,}000$
(\autoref{fig:closed-spectra} in \autoref{app:closure}). A nonlinear global constraint can
therefore be enforced on coarse data once its unclosed term is modelled, and a single scalar
is enough to restore the energy budget. However, matching the distribution of energy across scales
requires a closure with spectral structure.

%% file: figures/table1_body.tex
\begin{tabular}{@{}l ccccc ccccc@{}}
  \toprule
  & \multicolumn{5}{c}{E1 (resolved)} & \multicolumn{5}{c}{E2 (unresolved)} \\
  \cmidrule(lr){2-6} \cmidrule(lr){7-11}
  Model & step 64 & \shortstack{max\\$|\diver \vec{u}|$} & \shortstack{$|\langle \boldsymbol{u} \rangle_\Omega|$} & train & infer. & step 64 & \shortstack{max\\$|\diver \vec{u}|$} & \shortstack{$|\langle \boldsymbol{u} \rangle_\Omega|$} & train & infer. \\
  \midrule
  \shortstack[l]{FNO\\\strut} & \shortstack{$\mathbf{0.016}$\\$\pm0.002$} & \shortstack{$0.232$\\$\pm0.010$} & \shortstack{$3.3{\pm}1.3$\\$\times 10^{-4}$} & \shortstack{$1.00\times$\\\strut} & \shortstack{$1.00\times$\\\strut} & \shortstack{$1.11$\\$\pm0.01$} & \shortstack{$0.51$\\$\pm0.04$} & \shortstack{$7.5{\pm}2.6$\\$\times 10^{-4}$} & \shortstack{$1.00\times$\\\strut} & \shortstack{$1.00\times$\\\strut} \\
  \shortstack[l]{FNO+proj (cont)\\\strut} & \shortstack{$0.017$\\$\pm0.001$} & \shortstack{$\mathbf{2.6{\pm}0.01}$\\$\times 10^{-5}$} & \shortstack{$5.4{\pm}1.8$\\$\times 10^{-4}$} & \shortstack{$1.04\times$\\\strut} & \shortstack{$1.04\times$\\\strut} & \shortstack{$1.09$\\$\pm0.02$} & \shortstack{$5.0{\pm}0.1$\\$\times 10^{-7}$} & \shortstack{$1.1{\pm}0.2$\\$\times 10^{-3}$} & \shortstack{$1.01\times$\\\strut} & \shortstack{$1.04\times$\\\strut} \\
  \shortstack[l]{FNO+proj (+mom)\\\strut} & \shortstack{$0.018$\\$\pm0.001$} & \shortstack{$\mathbf{2.6{\pm}0.02}$\\$\times 10^{-5}$} & \shortstack{$2.0{\pm}0.1$\\$\times 10^{-8}$} & \shortstack{$1.04\times$\\\strut} & \shortstack{$1.07\times$\\\strut} & \shortstack{$1.08$\\$\pm0.01$} & \shortstack{$5.0{\pm}0.1$\\$\times 10^{-7}$} & \shortstack{$8.6{\pm}1.1$\\$\times 10^{-11}$} & \shortstack{$1.02\times$\\\strut} & \shortstack{$1.05\times$\\\strut} \\
  \shortstack[l]{FNO+proj (+energy)\\\strut} & \shortstack{$0.018$\\$\pm0.001$} & \shortstack{$\mathbf{2.6{\pm}0.01}$\\$\times 10^{-5}$} & \shortstack{$\mathbf{1.9{\pm}0.04}$\\$\times 10^{-8}$} & \shortstack{$1.49\times$\\\strut} & \shortstack{$1.34\times$\\\strut} & \shortstack{$\mathbf{1.01}$$^{\dagger}$\\$\pm0.01$} & \shortstack{$\mathbf{4.8{\pm}0.1}$\\$\times 10^{-7}$} & \shortstack{$\mathbf{7.7{\pm}0.6}$\\$\times 10^{-11}$} & \shortstack{$1.23\times$\\\strut} & \shortstack{$1.34\times$\\\strut} \\
  \midrule
  \shortstack[l]{PINO (cont only)\\\strut} & \shortstack{$0.026$\\$\pm0.004$} & \shortstack{$0.096$\\$\pm0.055$} & \shortstack{$1.1{\pm}1.1$\\$\times 10^{-3}$} & \shortstack{$1.10\times$\\\strut} & \shortstack{$1.01\times$\\\strut} & \shortstack{$1.09$\\$\pm0.01$} & \shortstack{$0.30$\\$\pm0.03$} & \shortstack{$9.0{\pm}4.6$\\$\times 10^{-4}$} & \shortstack{$1.04\times$\\\strut} & \shortstack{$0.96\times$\\\strut} \\
  \shortstack[l]{PINO (cont only) + TTO\\\strut} & \shortstack{$0.425$\\$\pm0.052$} & \shortstack{$8.0{\pm}1.3$\\$\times 10^{-3}$} & \shortstack{$0.11$\\$\pm0.06$} & \shortstack{$1.10\times$\\\strut} & \shortstack{$7312\times$\\\strut} & \shortstack{$1.05$$^{\dagger}$\\$\pm0.07$} & \shortstack{$5.0{\pm}0.4$\\$\times 10^{-3}$} & \shortstack{$7.5{\pm}4.5$\\$\times 10^{-3}$} & \shortstack{$1.04\times$\\\strut} & \shortstack{$1388\times$\\\strut} \\
  \shortstack[l]{PINO\\\strut} & \shortstack{$0.026$\\$\pm0.005$} & \shortstack{$0.217$\\$\pm0.142$} & \shortstack{$1.0{\pm}0.5$\\$\times 10^{-3}$} & \shortstack{$1.10\times$\\\strut} & \shortstack{$1.03\times$\\\strut} & \shortstack{$1.07$\\$\pm0.01$} & \shortstack{$0.17$\\$\pm0.02$} & \shortstack{$8.3{\pm}1.0$\\$\times 10^{-4}$} & \shortstack{$1.05\times$\\\strut} & \shortstack{$0.94\times$\\\strut} \\
  \shortstack[l]{PINO + TTO\\\strut} & \shortstack{$0.017$\\$\pm0.006$} & \shortstack{$0.024$\\$\pm0.010$} & \shortstack{$7.0{\pm}4.7$\\$\times 10^{-4}$} & \shortstack{$1.10\times$\\\strut} & \shortstack{$3535\times$\\\strut} & \shortstack{$1.02$$^{\dagger}$\\$\pm0.08$} & \shortstack{$2.8{\pm}0.2$\\$\times 10^{-3}$} & \shortstack{$2.1{\pm}1.0$\\$\times 10^{-4}$} & \shortstack{$1.05\times$\\\strut} & \shortstack{$2237\times$\\\strut} \\
  \bottomrule
  \multicolumn{11}{@{}p{\textwidth}@{}}{\footnotesize $^{\dagger}$These lower values reflect energy loss rather than closer dynamics (\autoref{sec:results-accuracy}).} \\
\end{tabular}

%% file: sections/06_discussion.tex
\section{Discussion}
\label{sec:discussion}

The introduction framed these experiments as a controlled proxy for neural operators that see
only coarse-grained observations, for example geophysical fields in weather and climate
prediction. E1 is the case where a residual loss is valid: the Kolmogorov flow is closed up to
discretisation error. E2 is the case where it is not, since the forcing lies entirely beyond the
cutoff and the energy in the resolved band arrives by nonlinear transfer from scales the
learner never sees. On the ground truth data the pointwise momentum residual is $0.094$ on E1
and $0.648$ on E2 (\autoref{tab:truth}). The comparisons in \autoref{sec:results} follow that
split, and they add two things the filtering argument alone does not provide: the form the
implicit closure takes in the learned operator, and the behaviour of penalised/constrained
models at horizons where the flow has decorrelated from its initial condition.

On E2, driving the unclosed residual towards zero does not recover the missing subgrid flux.
It removes energy from the wavenumber band the residual cannot account for, and the unclosed
energy balance projection does the same just more slowly. This is the optimiser-selected closure
described in the introduction: the data term and the physics term cannot be satisfied
together, so training selects a compromise, and the compromise is a lower-energy field.
Optimising the residual harder does not change the outcome. Test-time optimisation reduces
the residual by one to two orders of magnitude and never blows up, but only because it drains
the field: at step $64$ it holds $0.20$\% of its energy above wavenumber $16$
against the ground truth's $18.3$\%.

On E1 the equations close, so the residual is a valid constraint, yet enforcing it as a
penalty still does not help (\autoref{sec:results-accuracy}, \autoref{sec:results-stability}). A valid residual is still a penalty with a
coefficient that must be balanced against the data term; it is not equivalent to restricting
the model to the admissible set.

The constraint columns of \autoref{tab:main} measure what motivated enforcing constraints in
the first place: downstream models and analyses assume that the constraints are satisfied. On
both flows the projection maintains continuity and the domain-mean velocity constraints at the precision of
the stored data --- several orders of magnitude below the FNO and both PINOs --- for an
additional $2$--$4$\% of training time and $4$--$7$\% of inference time. This is a guarantee
that a penalty cannot provide: it holds at every training step and at every rollout step, not
on average over a training distribution. A divergence-free field can be passed to a tracer
solver or inverted for a streamfunction without spurious sources; an exactly conserved domain
mean cannot accumulate a drift over a long integration.

The closure result of \autoref{sec:results-probe} qualifies the filtering criterion: a
nonlinear global constraint is not invalid in itself, it is invalid until its unclosed term is
modelled. One fitted scalar keeps the energy bounded under the modelled balance over
$2{,}000$ steps. The fraction of energy above wavenumber $16$ still grows from the ground
truth's $0.18$ to $0.50$, so the model drifts away from the attractor of the flow despite
remaining within the energy band. Keeping a learned operator on the attractor is the problem the
Markov neural operator addresses \citep{li2022dissipative}, where a one-step operator is trained with
a Sobolev-norm loss, which weights prediction errors across scales, together with a
dissipativity regularisation that trains the operator to dissipate
energy on states far away from the attractor. Both parts compose directly with the projection layer used
here: the projection would enforce the linear constraints and the modelled energy balance,
while the Sobolev-norm term penalises errors in scale-by-scale transfer that a
single scalar cannot represent. The spectral drift also has an architectural component. The
FNO concentrates its error near the spectral cutoff, and an architecture with less spectral
bias, such as the multiscale wavelet transformer of \citet{wang2026mswt}, is a candidate
backbone for the same projection. We leave both combinations to future work.

\paragraph{Limitations.} The evidence is two flows, both two-dimensional, both on a periodic
domain, and one architecture. The filter is a sharp spectral cutoff, which is the cleanest
case for the argument: a smooth filter leaves a subgrid forcing that is small rather than
exactly zero, and these experiments do not locate the boundary of that regime. The stability
criterion is a band on the kinetic energy, which detects a blow-up or a drain but not a rollout
that stays bounded while ceasing to resemble the flow; the closed-energy model is exactly such a case,
and only the spectral diagnostics expose it. The linear constraint right-hand sides here are zero,
so the momentum balance constraint reduces to zeroing one Fourier mode. A non-zero, time-dependent target
would exercise more of \eqref{eq:projection}. The constant-flux closure assumes a
statistically stationary flow: it cannot represent state-dependent variation in
the subgrid flux, and its accuracy here rests on E2 being stationary.

%% file: sections/07_conclusion.tex
\section{Conclusion}
\label{sec:conclusion}

Linear constraints that commute with the filter survive coarse-graining; nonlinear terms generally require closure. A
neural operator trained on coarse-grained data can therefore enforce the
linear constraints exactly. On both experiments, we apply a closed-form spectral projection that enforces
continuity and the global momentum balance at the precision of the stored
data for an additional $2$--$4$\% of training time. Residual penalties added to the loss, either at
training time or at test time, provide no consistent accuracy benefit and
remove energy from the flow. A nonlinear constraint is not
invalid in itself, it is invalid until its unclosed term is modelled. A
constant subgrid flux fitted from the coarse training data restores the
global energy balance, allowing it to be enforced alongside the linear constraints. The
closed model maintains bounded energy under the modelled balance while still drifting from the
attractor, as the constraint does not preserve the distribution of energy across scales.

%% file: sections/02_filtering.tex
\section{The filtering criterion}
\label{sec:filtering}

For a state $u$ on $[0,T] \times \domain$, write the PDE problem governing the evolution of that state as
\begin{equation}
  \label{eq:constrained-pde}
  \pdeop[u] = f \quad\text{and}\quad \locop[u] = 0 \ \text{ in } \domain, \qquad
  \bcop[u] = g \ \text{ on } \partial\domain, \qquad
  \globop_{k}[u](t) = c_{k}(t), \quad k = 1,\dots,K ,
\end{equation}
with $\pdeop$ being the evolution operator, $\locop$ a set of constraints holding at every interior
point, $\bcop$ a boundary operator, and $\globop_{k}$ functionals of the whole field. Now, coarse-grain with a filter of width
$\delta$,
\begin{equation}
  \label{eq:filter}
  \filt{\phi}(x,t) = \int G_{\delta}(x-y)\,\phi(y,t)\,\mathrm{d}y ,
\end{equation}
with $\int G_{\delta}\,\mathrm{d}y = 1$. We use a fixed, translation-invariant filter on a
periodic domain, so it commutes with differentiation and preserves domain integrals. Split
the state into the resolved part $\filt{u}$ and the residual part $u' = u - \filt{u}$.

\paragraph{Linear constraints remain valid.} For a linear constraint operator that commutes
with \eqref{eq:filter}, $\locop[\filt{u}] = \filt{\locop[u]} = 0$. This holds for
constant-coefficient differential operators such as divergence; normalisation also preserves
unweighted domain integrals. Linearity alone does not establish commutation: variable
coefficients and boundary conditions require separate compatibility with the filter.
Under these assumptions, the constraint remains exact regardless of $\delta$, the shape of
$G_{\delta}$ or how much energy the filter removes.

\paragraph{The evolution equation does not.} Split $\pdeop$ into its linear and nonlinear
parts. Under the assumptions above, the linear part commutes with the filter. The nonlinear
part generally leaves behind $\filt{\pdeop_{\mathrm{nl}}[u]} - \pdeop_{\mathrm{nl}}[\filt{u}]$,
which depends on unresolved products and cannot in general be evaluated from the coarse
field alone, so representing it requires a closure model. A sharp spectral cutoff also
removes forcing supported entirely beyond it: the filtered equation then has no explicit
forcing term, while the flow it describes is forcing-driven.

\paragraph{A residual loss on filtered data is a closure model.} Write the filtered evolution
equation as $\pdeop[\filt{u}] = \filt{f} - \mathcal{R}[u,u']$, where $\mathcal{R}$ is the
unclosed residual. A physics-informed loss penalises
$\|\pdeop[\nnop_\parameters(\filt{u})] - \filt{f}\|^{2}$, and driving that penalty to zero
asserts $\mathcal{R} = 0$, i.e. it is a closure model that sets the subgrid term to zero. Explicit learned closures
\citep{guan2023physicsconstrained,zhao2025lesnets,wang2024coarsegraining,kochkov2021mlcfd,stachenfeld2022learned}
are approximations too, but they can be inspected, tested and reported. This asymmetry
motivates the paper: a commuting linear constraint is exact on filtered data at no modelling cost,
whereas a nonlinear residual on filtered data is a modelling choice, regardless of whether it is made
deliberately.

\paragraph{Example: the Navier--Stokes equations.} Consider the incompressible Navier--Stokes equations on a periodic domain, with viscosity
$\nu$ and a linear drag coefficient $\alpha$,
\begin{equation}
  \label{eq:ns}
  \diver \vec{u} = 0, \qquad
  \partial_{t}\vec{u} + \diver(\vec{u} \otimes \vec{u}) + \nabla p - \nu \Delta \vec{u}
  + \alpha \vec{u} = \vec{f} .
\end{equation}
Filtering the continuity equation returns $\diver \filt{\vec{u}} = 0$, because divergence
commutes with this filter. However, the filtered momentum equation does not close: filtering the convective term produces the subgrid stress
$\sgs = \filt{\vec{u} \otimes \vec{u}} - \filt{\vec{u}} \otimes \filt{\vec{u}}$, containing contributions from the
unresolved field, and the filtered equation becomes
\begin{equation}
  \label{eq:filtered-momentum}
  \partial_t \filt{\vec{u}} + \diver(\filt{\vec{u}} \otimes \filt{\vec{u}}) + \nabla \filt{p}
  - \nu \Delta \filt{\vec{u}} + \alpha \filt{\vec{u}} = \filt{\vec{f}} - \diver \sgs .
\end{equation}
A residual loss written on the coarse fields is
\eqref{eq:filtered-momentum} with the right-hand side truncated to $\filt{\vec{f}}$, so
minimising it asserts $\diver \sgs = 0$.

Integrating \eqref{eq:filtered-momentum} over the periodic domain removes every divergence, $\diver \sgs$ included, as per the divergence theorem, along with the pressure gradient and the viscous term. What is left is
the global momentum balance,
\begin{equation}
  \label{eq:momentum-balance}
  \frac{\mathrm{d}}{\mathrm{d}t} \int_{\domain} \filt{\vec{u}} \, \mathrm{d}V
  + \alpha \int_{\domain} \filt{\vec{u}} \, \mathrm{d}V
  = \int_{\domain} \filt{\vec{f}} \, \mathrm{d}V ,
\end{equation}
every term of which is linear in $\filt{\vec{u}}$. We enforce \eqref{eq:momentum-balance} and continuity as exact constraints via the projection described in \autoref{sec:projection}.

Conversely, contracting \eqref{eq:filtered-momentum} with
$\filt{\vec{u}}$ and integrating gives the global energy balance,
\begin{equation}
  \label{eq:filtered-energy}
  \frac{\mathrm{d}}{\mathrm{d}t} \int_{\domain} \tfrac{1}{2}|\filt{\vec{u}}|^{2}\,\mathrm{d}V
  = -\nu \int_{\domain} |\nabla \filt{\vec{u}}|^{2}\,\mathrm{d}V
    - 2\alpha \int_{\domain} \tfrac{1}{2}|\filt{\vec{u}}|^{2}\,\mathrm{d}V
    + \int_{\domain} \filt{\vec{u}} \cdot \filt{\vec{f}}\,\mathrm{d}V
    - \int_{\domain} \filt{\vec{u}} \cdot (\diver \sgs)\,\mathrm{d}V .
\end{equation}
The convective and pressure terms vanish after integration by parts because
$\diver\filt{\vec{u}} = 0$. The last term is the only one that does not close,
\begin{equation}
  \label{eq:sgs-flux}
  \Pi_{\delta}
  = -\int_{\domain} \filt{\vec{u}} \cdot (\diver \sgs) \, \mathrm{d}V,
\end{equation}
which is the net rate at which unresolved scales do work on the resolved field. For a sharp
radial cutoff it equals $\Pi(K)$ of \eqref{eq:shell-budget}. E2 uses a rectangular cutoff,
so the radial shell flux is a diagnostic comparison rather than the same integral.
On E2, omitting this positive energy supply from \eqref{eq:filtered-energy} forces the
resolved energy to decay. We
enforce this balance in two forms: unclosed, as a probe of what a mis-specified constraint
does to a model, and closed by the constant-flux model of \autoref{app:closure} described in
\autoref{sec:results-probe}.

%% file: sections/03_projection.tex
\section{Exact linear constraints by projection}
\label{sec:projection}

Once the admissible set is linear, the projection onto it has a closed form. Discretise the linear
constraints from \autoref{sec:filtering} as $\mat{A}\vec{u} = \vec{b}$, with
$\vec{u} \in \R^{n}$ the state on the grid and $\mat{A} \in \R^{m \times n}$ carrying one row
per constraint, $n$ being the total degrees of freedom on the grid and $m$ the number of discrete linear equalities. We seek the smallest correction to an unconstrained prediction
$\vec{u}^{\,*}$ that satisfies them, measured in the norm of a symmetric positive-definite mass
matrix $\mat{M}$ holding the grid weights (i.e. the discrete inner product whose weights turn a sum over grid values into an integral over the domain). The Karush--Kuhn--Tucker conditions of
$\min_{\vec{u}} \tfrac{1}{2} \| \vec{u} - \vec{u}^{\,*} \|_{\mat{M}}^{2}$ subject to
$\mat{A}\vec{u} = \vec{b}$ are linear. They are necessary and sufficient, because the problem
is a strictly convex quadratic with affine constraints. The Lagrangian is
$\tfrac{1}{2}(\vec{u}-\vec{u}^{\,*})^{\transpose}\mat{M}(\vec{u}-\vec{u}^{\,*})
+ \vec{\lambda}^{\transpose}(\mat{A}\vec{u}-\vec{b})$. Stationarity gives
$\vec{u} = \vec{u}^{\,*} - \mat{M}^{-1}\mat{A}^{\transpose}\vec{\lambda}$; substituting into
$\mat{A}\vec{u}=\vec{b}$ and eliminating $\vec{\lambda}$ returns
\begin{equation}
  \label{eq:projection}
  \vec{u}^{\,p}
  = \vec{u}^{\,*} - \mat{M}^{-1} \mat{A}^{\transpose}
    \left( \mat{A} \mat{M}^{-1} \mat{A}^{\transpose} \right)^{-1}
    \left( \mat{A} \vec{u}^{\,*} - \vec{b} \right) .
\end{equation}
The Schur complement $\mat{A}\mat{M}^{-1}\mat{A}^{\transpose}$ is invertible exactly when
$\mat{A}$ has full row rank. For the continuity rows
$\vec{b}$ is zero. For the momentum rows it is \eqref{eq:momentum-balance} integrated across
one step, so that $\vec{b}$ is the domain-mean momentum the coarse field is required to have
at the end of the step; on both flows used in this paper the forcing has no domain mean and the initial
mean is zero, so those rows set the mean velocity to zero.

The map $\proj : \vec{u}^{\,*} \mapsto \vec{u}^{\,p}$ has no parameters and is differentiable,
so $\proj \circ \nnop_{\parameters}$ trains end to end by ordinary backpropagation
\citep{chen2024kkthpinn,duruisseaux2024hard}, and appending it costs no expressivity (any
admissible field is a fixed point of $\proj$ \citep{min2024hardnet}). A single $\mat{A}$ contains every
row, so pointwise and global constraints are imposed jointly rather than in sequence,
imposing them one after the other would in general undo the first with the second. There is no
coefficient anywhere in \eqref{eq:projection}, which is the practical difference from a
penalty: the constraint is satisfied at every step of training, at every step of the rollout,
and at every point of the parameter space, including the initialisation.

\paragraph{Periodic grids.} In this case $\mat{M}$ is a positive multiple of the identity and therefore cancels from the projection formula, and both families of rows are diagonal in the Fourier transform. A divergence row acts as
$\mathrm{i}\vec{k} \cdot \hat{\vec{u}}(\vec{k})$ and couples only the components at one
wavevector, so \eqref{eq:projection} reduces mode by mode to the Leray projection
$(\mat{I} - \vec{k}\vec{k}^{\transpose}/|\vec{k}|^{2})\hat{\vec{u}}(\vec{k})$ commonly used in pseudo-spectral methods. A domain-mean
row is $|\domain|\,\hat{\vec{u}}(\vec{0})$ and touches only the zero wavevector, which the
Leray projection rows leave untouched. The two families therefore act on disjoint modes, the joint solve is
one pass over the spectrum, and the whole projection costs a forward and an inverse transform,
$O(N \log N)$ in the number of grid points, with no linear system to assemble. The projection
layer and its spectral form appear in prior work: \eqref{eq:projection} is the hard-constraint
output layer of \citet{chen2024kkthpinn} and \citet{min2024hardnet}, treated in generality by
\citet{grontas2026pinet}, and its Leray form appended to an FNO and trained end to end on
two-dimensional Navier--Stokes appears in
\citet{duruisseaux2024hard,xu2025pcno,li2026projectgenerate}. What this paper adds is the
zero-wavevector rows in the same $\mat{A}$.

\paragraph{The energy balance.} The global energy balance is a single scalar equation,
quadratic in $\vec{u}$, so it has no row in $\mat{A}$. We impose it after \eqref{eq:projection}
on the discrete trapezoidal balance between consecutive steps,
\begin{equation}
  \label{eq:energy-balance}
  h(\vec{u}) = E_{n+1} - E_{n}
  + \tfrac{\Delta t}{2}\left[ \nu (Z_{n} + Z_{n+1}) + 2\alpha (E_{n} + E_{n+1})
  - (P_{n} + P_{n+1}) \right] = 0 ,
\end{equation}
with $E = \tfrac{1}{2}\int_{\domain} |\vec{u}|^{2}$, $Z = \int_{\domain} |\nabla \vec{u}|^{2}$ and
$P = \int_{\domain} \vec{u} \cdot \vec{f}$, by at most $20$ minimum-norm steps
$\vec{u} \leftarrow \vec{u} - h(\vec{u})\,\nabla h(\vec{u})/\|\nabla h(\vec{u})\|^{2}$ in Fourier
space, with residual tolerance $10^{-10}$. The step direction is
$(1 + \Delta t\,\alpha + \Delta t\,\nu |\vec{k}|^{2})\hat{\vec{u}}(\vec{k})
- \tfrac{\Delta t}{2}\hat{\vec{f}}(\vec{k})$, a per-mode multiple of the field plus the forcing,
so it is divergence-free wherever the field and the forcing are and vanishes at
$\vec{k} = \vec{0}$ whenever the domain means do. Unclosed, the step is an instrument for
measuring what enforcing a non-closing constraint does to a model.

Equation \eqref{eq:projection} is the finite-dimensional case of a constrained
reproducing-kernel construction \citep{jidling2017linearly,langehegermann2018algorithmic};
non-periodic geometries and non-uniform meshes are future work.

%% file: sections/04_setup.tex
\section{Experiments}
\label{sec:setup}
\label{app:setup}

The filtering criterion of \autoref{sec:filtering} is a statement about the data, so it can
be checked before any model exists. We use it to construct two flows, one resolved and one
unresolved, both observed on the same $64^{2}$ grid.

\paragraph{E1, the resolved control.} This is the Kolmogorov-flow benchmark of
\citet{li2021pino}: Navier--Stokes at $\mathrm{Re} = 500$ on the torus $(0,2\pi)^{2}$, forced
by
$\vec{f} = \sin(4x_{2})\,\vec{e}_{1}$, with $4{,}000$ trajectories of $65$ snapshots. It is
published in vorticity form; we recover velocity and pressure by spectral inversion of the
Poisson equations. The reconstructed velocity is divergence-free by construction, but the
unconstrained model's velocity predictions need not satisfy this or the global momentum balance. The snapshot
spacing is $\Delta t = 1/64$. The data are used at their native $64^{2}$ resolution. The
forcing, the energy-containing scales and the dissipation all sit inside the grid, and the
equation closes up to discretisation error. This makes E1 the case where a residual loss is
valid to enforce.

\paragraph{E2, the unresolved case.} We apply the Dedalus pseudo-spectral solver
\citep{burns2020dedalus} to stochastically forced two-dimensional turbulence
\citep{boffetta2012twodimensional} on the periodic domain $(0,2\pi)^2$. Five realisations are integrated
to $t = 2000$ at $1024^{2}$ with $\nu = 9.3 \times 10^{-6}$ and linear drag
$\alpha = 3 \times 10^{-3}$, under an Ornstein--Uhlenbeck forcing confined to
$|\vec{k}| \in [56, 72]$ with $\varepsilon = 1.71\times10^{-5}$ and correlation time
$\tau = 0.3$, at a solver step of $8\times10^{-3}$. A spin-up phase is conducted up to $t = 924.8$; data prior to this is discarded. Fields are written at $64^{2}$ every
$0.08$ time units by a sharp spectral cutoff applied at $|k_{x}|, |k_{y}| \le 31$, placed
\emph{below} the forcing band. We retain every fifth snapshot, giving a model step of
$\Delta t = 0.4$. On the coarse field the forcing RMS is $5.9\times10^{-18}$ of
the velocity RMS. The coarse dynamics in the resolved band are therefore driven entirely by
nonlinear transfer from scales below the grid. E2 is the case where a residual loss is not
valid to enforce without an explicit closure.

The two flows therefore differ in how the resolved band is supplied with energy. Write $E(k)$
for the kinetic energy in the shell of wavenumber magnitude $k$, $T(k)$ for the net nonlinear
transfer out of that shell, and $F(k)$ for the energy the forcing injects. The shell budget is
\begin{equation}
  \label{eq:shell-budget}
  \frac{\mathrm{d}E(k)}{\mathrm{d}t} = -T(k) - 2(\nu k^{2} + \alpha)\,E(k) + F(k) .
\end{equation}
Summing over $|\vec{k}| \le K$ for a statistically stationary flow with no forcing inside the
band gives $\Pi(K) = D_{\mathrm{in}}(K)$, with $\Pi(K) = -\sum_{|\vec{k}| \le K} T(k)$ the
cumulative nonlinear transfer into the band and
$D_{\mathrm{in}}(K) = \sum_{|\vec{k}| \le K} 2(\nu k^{2} + \alpha)\,E(k)$ the dissipation
inside it.

\begin{figure}[t]
  \centering
  \includegraphics[width=1.0\linewidth]{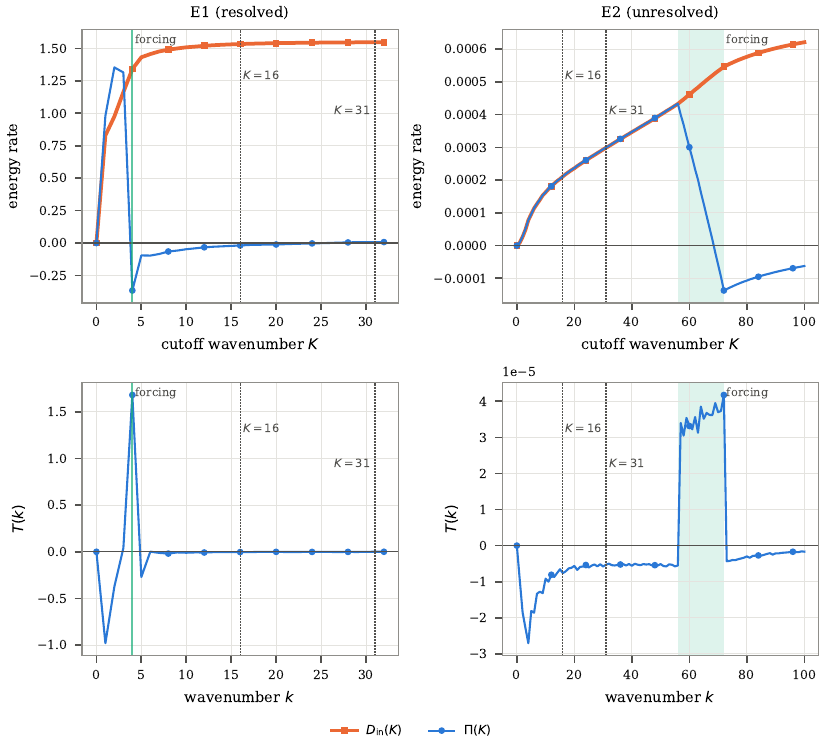}
  \caption{The energy-budget diagnostic of \eqref{eq:shell-budget}. Top: cumulative
  nonlinear transfer into the band at wavenumber $K$, $\Pi(K)$, as well as cumulative dissipation,
  $D_{\mathrm{in}}(K)$. Bottom: nonlinear transfer $T(k) = -\mathrm{d}\Pi/\mathrm{d}k$. Forcing is
  marked ($k = 4$ on E1; $|\vec{k}| \in [56,72]$ on E2) by the green shading.}
  \label{fig:flux}
\end{figure}

\autoref{fig:flux} plots both quantities. On E2 they coincide throughout the
resolved wavenumber band: measured on $1{,}100$ stationary snapshots of one realisation,
$\Pi/D_{\mathrm{in}}$ is $1.002$ at $K = 16$ and $1.003$ at $K = 31$. The resolved range is a
net receiver of energy from beyond the cutoff, and there is no resolved forcing to replace what
it dissipates. On E1 they do not coincide. The forcing sits at $k = 4$, and
the ratio $\Pi/D_{\mathrm{in}}$ is $-0.012$ at $K = 16$ and $+0.005$ at $K = 31$.

\autoref{tab:truth} reports the residuals of each constraint on the ground truth. On both
flows the two linear constraints hold to numerical round-off. On E1 the pointwise momentum
equation very nearly holds as well. On E2, both this residual and the global energy-balance
residual are $O(1)$ after normalisation.

\begin{table}[t]
  \caption{Ground-truth constraint diagnostics. The linear rows are absolute errors;
  the nonlinear rows are relative residuals, with normalisations defined below.}
  \label{tab:truth}
  \centering
  \small
  \begin{tabular}{llcc}
    \toprule
    & & E1 (resolved) & E2 (unresolved) \\
    \midrule
    \multirow{2}{*}{linear}
      & continuity, $\max|\diver \vec{u}|$        & $2.1\times10^{-5}$ & $4.2\times10^{-7}$ \\
      & momentum balance, $|\langle\vec{u}\rangle_\Omega|$ & $1.4\times10^{-8}$ & $5.4\times10^{-11}$ \\
    \addlinespace
    \multirow{2}{*}{nonlinear}
      & pointwise momentum residual                & $0.094$            & $0.648$ \\
      & global energy balance                      & $0.256$                & $0.935$ \\
    \bottomrule
  \end{tabular}
\end{table}

The pointwise momentum residual is normalised by the root-mean-square one-step velocity
increment. The relative energy residual is the mean over pairs of
\begin{equation}
  \label{eq:relative-energy-residual}
  \frac{|h|}{|E_{n+1}-E_n| + \tfrac{\Delta t}{2}
  [\nu(Z_n+Z_{n+1})+2\alpha(E_n+E_{n+1})+|P_n|+|P_{n+1}|]} ,
\end{equation}
with $h$, $E$, $Z$ and $P$ defined in \eqref{eq:energy-balance}.

\begin{figure}[t]
  \centering
  \includegraphics[width=1.0\linewidth]{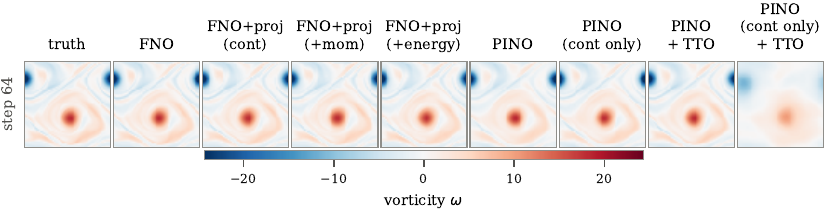}
  \caption{Vorticity at the final step of a $64$-step rollout on E1.}
  \label{fig:snapshots-e1}
\end{figure}

\begin{figure}[t]
  \centering
  \includegraphics[width=1.0\linewidth]{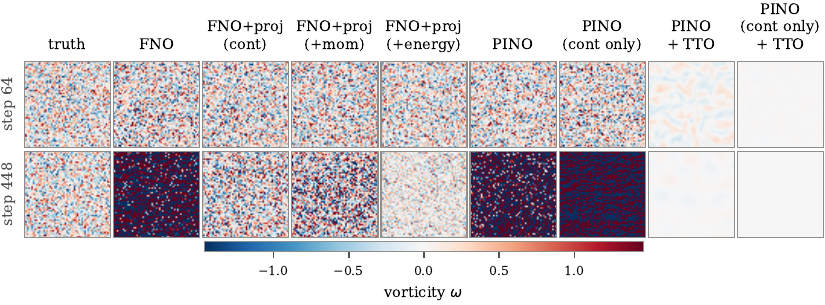}
  \caption{Vorticity at step $64$ of the standard rollout (top) and step $448$ of the extended rollout (bottom) on E2.}
  \label{fig:snapshots-e2}
\end{figure}

\autoref{fig:snapshots-e1} and \autoref{fig:snapshots-e2} show the vorticity field on one test
trajectory of each flow.

\paragraph{Architecture and training.} Every method on both flows uses the same Fourier neural
operator architecture: four spectral layers of width $64$, GELU activations, a
$128$-unit channel-mixing head, and increment prediction
$\vec{u}_{n+1} = \vec{u}_{n} + f_{\parameters}(\vec{u}_{n}, \text{grid})$, with the
projection applied to the predicted state $\vec{u}_{n+1}$ rather than the increment $f_{\parameters}(\vec{u}_{n}, \text{grid})$. The input is seven channels ($u$, $v$, $p$ and
four periodic grid features) and the output is three ($u$, $v$, $p$). The data loss is relative $L_{2}$, i.e. $\|\hat{u} - u\|_{2} / \|u\|_{2}$ over the flattened $(u,v,p)$ state. E1 retains $16$ modes per axis and E2
retains $32$. The Adam optimiser is used at a base learning rate of $10^{-3}$, weight decay $10^{-4}$, gradient-norm
clipping at $1.0$ and batch size $100$. E1 trains for $280$ epochs, with the learning rate
halved at epochs $112$, $196$ and $252$; E2 trains for $100$ epochs, with halvings at $40$,
$70$ and $90$. Each configuration uses seeds $1$--$5$ for initialisation, minibatch shuffling
and input noise. Training uses one-step pairs, without an additional initial-condition loss.

The data term has unit weight. For full PINO, continuity has weight $0.0767$ and each
momentum component has weight $0.0051$ on both flows. These weights were selected by
one-step validation error in one-seed sweeps with factor-of-three spacing, over $50$ epochs
on E1 and the full $100$-epoch schedule on E2. Continuity-only PINO uses $0.12$ on E1, selected for
rollout stability, and $0.026$ on E2, selected by one-step validation error. These weights
were chosen under the initial $16$-mode protocol and retained for the final experiments.

\paragraph{Protocol.}
\label{app:protocol}

Eight methods are compared on both flows, all sharing the one architecture, optimiser and
schedule described above: a plain FNO \citep{li2021fno}; three adding the projection of
\autoref{sec:projection} --- continuity alone, continuity with the momentum balance, and
both with the energy balance; a PINO with the continuity residual and both momentum
components; a PINO with the continuity residual only; and each of the two PINO methods
again with test-time optimisation (TTO), in which the trained weights are further optimised
on each test trajectory against its own PDE residual with no data term
(\autoref{app:tto}). We name the methods \emph{FNO},
\emph{FNO + proj (cont)}, \emph{FNO + proj (+ mom)}, \emph{FNO + proj (+ energy)},
\emph{PINO}, \emph{PINO (cont only)}, \emph{PINO + TTO} and
\emph{PINO (cont only) + TTO}.

Numerical summaries use seed means and sample standard deviations. Accuracy differences
are paired by seed and expressed as percentages of the FNO's mean error across seeds.
Energy error is $(E_{\mathrm{pred}}-E_{\mathrm{truth}})/E_{\mathrm{truth}}$.

\begin{figure}[t]
  \centering
  \includegraphics[width=1.0\linewidth]{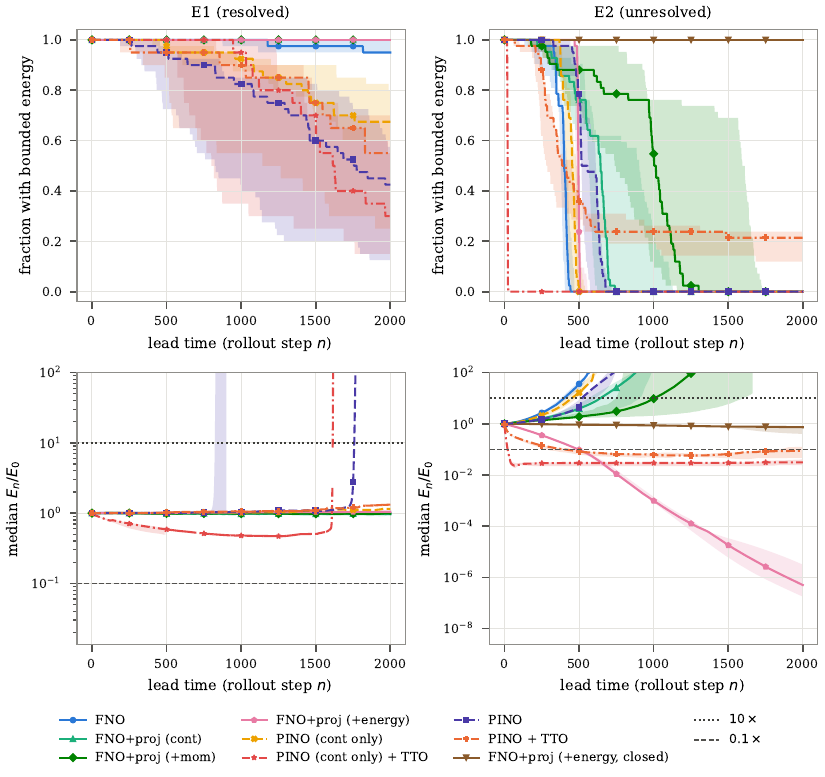}
  \caption{Free-running stability over $2{,}000$ steps, five seeds of $40$ rollouts on E1 ($20$ for the TTO methods)
  and $42$ on E2. Top: fraction that have stayed within $[0.1,10]$ times their initial energy.
  Bottom: median energy ratio $E_n/E_0$ within each seed. Curves and shading show the median
  and interquartile range across seeds; dotted lines mark the energy bounds.}
  \label{fig:stability}
\end{figure}

The standard evaluation rollout is $64$ steps on both flows. E1's released trajectories are
$65$ snapshots long, so this covers the entire available trajectory. E2's $2{,}688$ test
pairs give $42$ non-overlapping $64$-step windows; a separate extended evaluation uses six
$448$-step windows. Free-running evaluation uses $40$ initial conditions per seed on E1
($20$ for TTO) and $42$ on E2, each for $2{,}000$ steps. Across five seeds, this gives $200$
trajectories per method on E1 ($100$ for TTO) and $210$ on E2.
\autoref{fig:stability} records those free-running trajectories. Every method on E2 stays below the
$10\times$ bound for the first $128$ free-running steps, so no standard $64$-step evaluation
is cut short by a blow-up.

Separate train, validation and test splits are employed. Model training uses the training split
only. On E1, checkpoint selection is restricted to epochs at the full input-noise amplitude,
which under the ramp below is the final epoch. On E2, validation is evaluated every epoch and
the checkpoint with the lowest one-step data loss is selected. The test split is used to
report final results and is not used for checkpoint selection. On E1 the split is
$3{,}600/200/200$ trajectories. On E2 it is by realisation: train uses three seeds, validation
another, and the final seed is used for testing, so that a model is tested on a realisation
it has never seen. The split seeds and data preparation are specified in
\autoref{app:reproduction}.

\paragraph{Rollout stabilisers.} The two flows make use of two different methods for stabilising the autoregressive rollout. E1 trains with Gaussian input noise at
$\sigma = 2\times10^{-2}$ of the state, held at zero for the first fifth of the schedule and
then ramped linearly to full amplitude. E2 trains with an additional vorticity data term,
$\hat{\omega} = \mathrm{i}(k_{x}\hat{v} - k_{y}\hat{u})$, measured as a relative $L_{2}$
against the truth. This helps stability by penalising errors at higher wavenumbers more strongly, since taking the curl multiplies each mode by its wavenumber.
Its weight is $1.74$ for every E2 method, selected in a stabilisation screen and equal to
three times the weight that matches the vorticity and data terms at initialisation.

%% file: appendix/I_closure.tex
\section{Closing the energy balance with a constant flux}
\label{app:closure}

\subsection{The closure}

The global energy balance does not close on E2 because it omits the nonlinear energy flux
across the cutoff, $\Pi_{\delta}$ of \eqref{eq:sgs-flux}. We approximate this flux by a
constant, $\Pi_{\delta}(t) \approx \bar{S}$, giving the closed discrete balance
\begin{equation}
  \label{eq:closed-balance}
  h_{\bar{S}}(\vec{u}) = h(\vec{u}) - \Delta t\,\bar{S} = 0 .
\end{equation}
Here $h$ is the unclosed balance in \eqref{eq:energy-balance}. The constant shifts the target
of the constraint without changing its gradient, so we can use the Gauss--Newton update of
\autoref{sec:projection}. The update continues to preserve continuity and global momentum
balance. The only difference from the unclosed projection is the scalar $\bar{S}$.

We fit $\bar{S}$ from the coarse training data alone. Evaluated on a pair of ground-truth
snapshots, the residual of the unclosed balance equals the time integral of $\Pi_{\delta}$
over that step, up to the trapezoidal-rule error. This follows from the exact filtered energy
balance, \eqref{eq:filtered-energy}, which includes the subgrid flux. We therefore estimate
$\bar{S}$ as the mean of $h/\Delta t$ over the $6{,}885$ E2 training pairs. The fitted value is
$\bar{S} = 3.198\times10^{-4}$ per unit time, or
$\bar{S}\,\Delta t = 1.279\times10^{-4}$ per step. The standard error of this per-step
source is $1.4\times10^{-7}$, computed as the residual standard deviation divided by
$\sqrt{6{,}885}$, without adjustment for temporal correlation.

Two checks support this estimate. The radial shell flux at $K=31$ is
$2.99\times10^{-4}$ per unit time, within $7$\% of the fitted value, despite the difference
between the radial diagnostic and the rectangular data filter. On the held-out
validation split, the mean closed residual is $1.7\times10^{-8}$ per step, or $0.2$\% of the
ground truth's mean absolute one-step energy change of $9.2\times10^{-6}$. Applying the same fit to E1
gives a per-step source whose magnitude is $3.0$\% of its mean absolute one-step energy change, compared with
$1{,}367$\% on E2. This is consistent with E1 being resolved, so we do not train a separate
model with an energy closure on that flow.

\subsection{Results}

We add $\bar{S}$ to the FNO + proj (+ energy) configuration and train it end to end with five
seeds under the same protocol as the other methods (\autoref{app:protocol}). Its accuracy and
satisfaction of the two linear constraints are comparable to those of the other projected
models: relative $L_2$ error is $0.0571$ at step $1$,
$\max|\diver \vec{u}| = 5.0\times10^{-7}$ and $|\langle \boldsymbol{u} \rangle| =
7.8\times10^{-11}$. The closure substantially reduces the energy deficit. At step $64$, the
closed model's energy error relative to the ground truth is $-0.2 \pm 0.1$\%, compared with a
deficit of $20.9 \pm 0.1$\% for the unclosed projection. At step $448$, the respective errors
are $-8.6 \pm 4.0$\% and $-86.9 \pm 1.9$\%.

Over $2{,}000$ free-running steps ($210$ trajectories), the closed model's median energy is
$0.92$ of its initial value at step $448$ and $0.71$ at step $2{,}000$. Every trajectory ends
within $[0.1, 10]$ times its initial energy, making it the only E2 model for which all
trajectories satisfy this criterion. The unclosed projection's median energy is
$4.9\times10^{-7}$ of its initial value at step $2{,}000$. For the momentum-balance projection,
the median over the $210$ trajectories exceeds the upper limit of the $[0.1, 10]$ band at step
$975$; for the continuity-only projection, it exceeds that limit at step $584$.

\begin{figure}[h]
  \centering
  \includegraphics[width=1.0\linewidth]{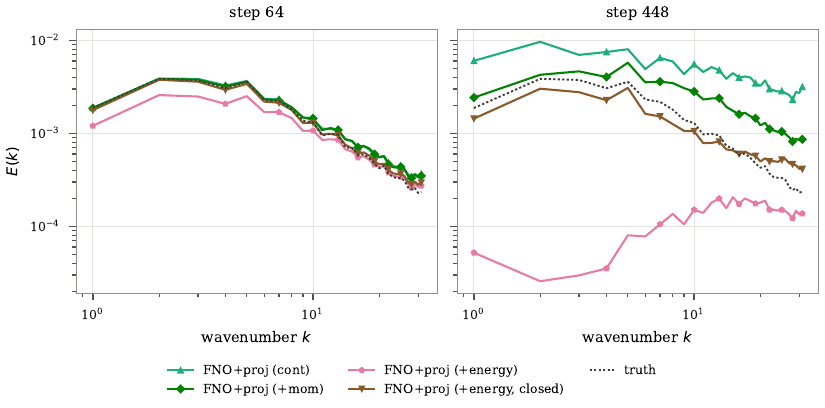}
  \caption{E2 energy spectra at the final step of the $64$-step rollout (left) and the
  $448$-step windows with ground truth (right), for the three projected models and the model
  with the closed energy balance. Curves on the left show the median over five seeds, each
  averaged over its test trajectories; those on the right show the median over the thirty
  seed--window rollouts. The dotted reference is the ground-truth spectrum at step $64$ on the
  left and its stationary mean over the $2{,}688$ test snapshots on the right. %
  }
  \label{fig:closed-spectra}
\end{figure}

\subsection{Distribution of energy across scales}

Although the closure restores the global energy budget, the predicted distribution of energy
across scales continues to depart from the ground truth (\autoref{fig:closed-spectra}). The
fraction of energy above wavenumber $16$ is $0.20$ at step $64$, $0.28$ at step $448$ and $0.50$
at step $2{,}000$, compared with the ground truth's $0.18$. By step $256$, the predicted vorticity range is $3.8\times$ that of the ground truth. The
constraint fixes the total energy budget but does not constrain how energy is distributed
over wavenumbers. Correcting the transfer between scales requires a closure with spectral
structure, which is beyond the constant-flux approximation considered here.

%% file: appendix/E_tto.tex
\section{Test-time optimisation}
\label{app:tto}

Test-time optimisation (TTO) further minimises the PDE residual on each test trajectory by
adapting the trained operator's parameters. We compare its constraint satisfaction and
computational cost with those of the projection. Unlike a penalty applied only during
training, TTO requires a separate optimisation for every trajectory at inference.

\subsection{Protocol}

Starting from the trained PINO checkpoint, we optimise all parameters of the operator
separately for each test trajectory, using the PDE residual evaluated over the adaptation
horizon. The residual terms retain the weights used to train the checkpoint:
continuity at $0.0767$ and each momentum component at $0.0051$ for the full PINO on both flows,
and continuity at $0.12$ on E1 and $0.026$ on E2 for the continuity-only methods. No data term
is used, since adaptation does not have access to the ground-truth trajectory.

\begin{description}
  \setlength{\itemsep}{2pt}
  \item[Optimiser] Adam with a learning rate of $3\times10^{-4}$, halved at iterations $100$
  and $160$, gradient-norm clipping at $1.0$, and a maximum of $201$ iterations.
  \item[Stopping] Optimisation stops after $30$ iterations without improvement. The weights
  with the lowest loss are restored at the end.
  \item[Guard] If a trajectory's loss exceeds ten times its lowest value so far, optimisation
  stops and the weights with the lowest finite loss are restored.
  \item[Evaluation] $20$ test trajectories per seed on E1; $42$ $64$-step windows and,
  separately, six $448$-step windows per seed on E2. Both flows use five seeds.
\end{description}

\paragraph{Loss guard.} For full-residual TTO on E1, $24$ of the $100$ trajectories
triggered the guard at the second iteration: the first optimiser step increased the loss to
more than ten times its initial value. In each case, the lowest loss was that of the original
checkpoint, so its weights were restored. These $24$ trajectories are therefore evaluated as
unadapted PINO rollouts in the TTO results. Of the remaining $76$ trajectories, $42$ reached
the iteration limit and $34$ stopped early. The reported means consequently include both
adapted and unadapted trajectories, which should be considered when interpreting the accuracy
comparison below.

\paragraph{Computational cost.} We report the wall-clock time required to produce one test
trajectory on one GPU, including both optimisation and the final rollout. Timing follows the
same convention as for the methods without adaptation. It excludes training the initial
checkpoint, whose cost is reported in the PINO row of \autoref{tab:main}.

\subsection{Results on E1}

On E1, TTO requires $66 \pm 18$ seconds per trajectory, compared with $0.020 \pm 0.001$ for
the projected model and $0.019 \pm 0.002$ for a plain FNO. This is a factor of $3{,}300$
relative to the projection. At step $1$, its relative $L_2$ error is $10.1 \pm 12.6$\% lower
than the FNO's, with an improvement on four of five seeds ($1.48\times10^{-3}$ compared with
$1.64\times10^{-3}$). The mean improvement is smaller than the seed standard deviation, and
at step $64$ the error is $7.6$\% higher than the FNO's.

TTO reduces the constraint residuals but does not enforce them exactly. The maximum divergence
$\max|\diver \vec{u}|$ is $2.4\times10^{-2}$ and the continuity residual is
$2.2\times10^{-3}$, approximately three orders of magnitude above the projection's
$2.6\times10^{-5}$ and $3.6\times10^{-6}$. The projection achieves these values with an
additional $4$\% of training time and a small inference overhead.

\subsection{Results on E2}

For the extended E2 evaluation, adaptation uses the first $64$ steps of each window, and the adapted
weights are then used for the full $448$-step rollout. This limits the cost of differentiating
through the rollout, which would otherwise span $448$ steps for every test trajectory. On
E1, the same adaptation horizon covers the full available trajectory. This extended evaluation
comprises six windows per seed, five seeds and two methods, giving sixty trajectories.
None triggered the loss guard. Unlike E1, where the initial checkpoint already has a low
residual, the unadapted E2 models produce divergent rollouts over this horizon.

\begin{center}
  \footnotesize
  \begin{tabular}{@{}lrrrr@{}}
    \toprule
    over the six $448$-step windows & PINO & PINO + TTO & PINO (cont) & PINO (cont) + TTO \\
    \midrule
    relative $L_{2}$, horizon mean   & $280$      & $1.12$               & $338$      & $1.06$ \\
    relative $L_{2}$ at step $64$    & $1.060$    & $1.029$              & $1.089$    & $1.049$ \\
    $\max|\diver \vec{u}|$          & $5.8\times10^{3}$ & $2.3\times10^{-2}$ & $1.1\times10^{4}$ & $4.9\times10^{-3}$ \\
    domain-mean velocity, final     & $69$       & $1.6\times10^{-3}$   & $111$      & $1.1\times10^{-2}$ \\
    continuity residual             & $63$       & $1.6\times10^{-2}$   & $92$       & $2.7\times10^{-3}$ \\
    seconds per trajectory          & $0.61$     & $120$                & $0.57$     & $79$ \\
    \bottomrule
  \end{tabular}
\end{center}

Adaptation prevents blow-up over this horizon and reduces the constraint diagnostics by four
to nine orders of magnitude. These errors remain above those of the projection, whose maximum
divergence and final domain-mean velocity are $4.5\times10^{-6}$ and $4.2\times10^{-10}$ on the
same six windows, at $0.70$ seconds per trajectory. All reported accuracy and constraint
diagnostics improve under adaptation in this comparison, but both unadapted PINO models have
diverged by step $448$. The improvements therefore reflect suppression of that divergence.
Both TTO methods lose energy, taking most trajectories below the $0.1\times$ bound
(\autoref{sec:results-stability}).

\subsection{Adapting against continuity alone}
\label{app:tto-cont}

On E1, adapting against continuity alone reduces the objective while degrading prediction
accuracy. The objective uses the continuity residual with its training weight of $0.12$ and
zero weight on the momentum terms.

\begin{center}
  \footnotesize
  \begin{tabular}{@{}lrrr@{}}
    \toprule
    E1, final step of the $64$-step rollout & before adaptation & after & PINO + TTO \\
    \midrule
    continuity residual (the objective) & $2.0\times10^{-3}$ & $9.0\times10^{-4}$ & $2.2\times10^{-3}$ \\
    $\max|\diver \vec{u}|$              & $1.6\times10^{-2}$ & $8.0\times10^{-3}$ & $2.4\times10^{-2}$ \\
    domain-mean velocity                & $1.1\times10^{-3}$ & $1.1\times10^{-1}$ & $7.0\times10^{-4}$ \\
    relative $L_{2}$                    & $0.023$            & $0.425 \pm 0.052$  & $0.017$ \\
    seconds per trajectory              & ---                & $137$              & $66$ \\
    \bottomrule
  \end{tabular}
\end{center}

This loss of accuracy occurs on all five seeds and is accompanied by a large increase in the
domain-mean velocity. A spatially uniform velocity field has zero divergence, so the
$\vec{k} = \vec{0}$ mode lies in the kernel of the divergence operator. A continuity-only
objective therefore does not penalise changes to the domain mean. Over the course of the rollout, the domain-mean velocity increases by two orders of magnitude.
The momentum residual, which is absent from the objective, rises from $0.093$ (compared with
$0.094$ for the ground truth) to $0.72$. When the objective includes continuity and both
momentum components, adaptation reduces the domain-mean velocity and the momentum residual
falls from $0.092$ to $0.064$.

Continuity satisfies the filtering criterion of \autoref{sec:filtering}, but enforcing a
valid constraint alone does not prevent drift in quantities that it leaves unconstrained.
Here, continuity does not determine the $\vec{k} = \vec{0}$ mode. The projection addresses
this by enforcing continuity and global momentum balance jointly in \eqref{eq:projection}:
both sets of constraints are included in $\mat{A}$, and the momentum rows fix the domain mean
left unconstrained by the divergence rows.

%% file: appendix/G_reproduction.tex
\section{Reproducibility}
\label{app:reproduction}

\paragraph{Availability.} The code that produced every result in this paper is released at \url{https://github.com/m-groom/physics-constrained-neural-operators}.
The E2 dataset, along with its normalisation statistics, is
available on Zenodo at \url{https://doi.org/10.5281/zenodo.23113478}; the
solver configuration that generated it is at \url{https://github.com/m-groom/NS2D}.
The E1 dataset is available
from the \href{https://github.com/neuraloperator/physics_informed#data-description}{PINO
repository} of \citet{li2021pino}. The E2 simulations use
\href{https://github.com/DedalusProject/dedalus}{Dedalus} \citep{burns2020dedalus}, version
$3.0.4$, with the equations and parameters in \autoref{sec:setup}.

\paragraph{Data preparation.} E1 uses \texttt{NS\_fft\_Re500\_T4000.npy}. Before spectral
inversion to velocity and pressure, we remove the mean vorticity and, as a numerical
convention, the Nyquist row and column to avoid ambiguity in the spectral derivatives. The pressure has zero domain
mean. A NumPy random permutation with seed $1234$ assigns the first $200$ trajectories to
validation, the next $200$ to testing and the remaining $3{,}600$ to training.

For E2, simulation seeds $1234$--$1236$ supply training data, $1237$ validation and $1238$
testing. After the spin-up cut and temporal stride in \autoref{sec:setup}, each realisation
contains $2{,}688$ consecutive pairs. From each training realisation, the final $384$ pairs
are reserved as a secondary temporal test set, separated from training by eight strided
snapshots. This leaves $2{,}295$ pairs per training realisation, or $6{,}885$ in total; the
reported E2 results use the separate test realisation. Normalisation statistics are computed
from training inputs only: each channel is centred on its mean, the velocity components
share a standard deviation, and pressure has its own. The projection and physical residuals
are evaluated after returning the fields to physical units.

\paragraph{Software.} Training and evaluation use Python $3.12$, PyTorch
$2.5.1$ and CUDA $12.2$, with single-precision model parameters and fields. %

\paragraph{Hardware and cost.} Each training or evaluation run uses one NVIDIA H100 GPU
with $94$\,GB of device memory. An FNO training run takes approximately six hours on E1
($280$ epochs at about $80$ seconds per epoch) and ten minutes on E2 ($100$ epochs at about
$6$ seconds per epoch). \autoref{tab:main} gives the cost ratios for the other methods, and
\autoref{app:tto} gives the absolute TTO times. The five E2 simulations each used $32$ CPU
ranks for $9$--$10$ hours and produced approximately $22$\,GB of output. Preparing the
coarse dataset took $15$ minutes on eight CPU cores with a peak host-memory use of
$39$\,GB. These costs exclude preliminary experiments and hyperparameter screens.

\paragraph{Asset licences.} Dedalus is distributed under the GNU General Public License
version $3$, and the PINO repository under the Apache License $2.0$. The E1 data are obtained
from the original authors' download linked above; no separate dataset licence is specified
there.